\documentstyle[12pt,epsf]{article}

\begin{document}
\begin{flushright}
IFUM 593/FT \\
IFUP-TH 35/97 \\
UCY-PHY-97/07
\end{flushright}
\vskip 2cm

\centerline{{\bf Asymptotic scaling corrections in $QCD$
with Wilson fermions }}
\centerline{{\bf from the 3-loop average plaquette}}
\vskip 5mm
\centerline{B. All\'es$^{\rm a}$, A. Feo$^{\rm b}$, 
H. Panagopoulos$^{\rm c}$}
\vskip 5mm
\centerline{$^{\rm a}${\it Dipartimento di Fisica, Sezione Teorica, 
Universit\`a di Milano and INFN,}} 
\centerline{\it Via Celoria 16, 20133-Milano, Italy}
\vskip 3mm
\centerline{$^{\rm b}${\it Dipartimento di Fisica dell'Universit\`a,}} 
\centerline{\it Piazza Torricelli 2, 56126-Pisa, Italy}
\vskip 3mm
\centerline{$^{\rm c}${\it Department of Natural Sciences, 
University of Cyprus, }}
\centerline{\it P.O. Box 537, Nicosia CY-1678, Cyprus}

\vskip 2cm

\begin{abstract}
We calculate the 3-loop perturbative expansion of the average plaquette 
in lattice QCD with $N_f$ massive Wilson fermions and gauge group $SU(N)$.
The corrections to asymptotic scaling in the corresponding energy scheme
are also evaluated. We have also improved the accuracy of the already
known pure gluonic results at 2 and 3 loops.
\end{abstract}

\vfill\eject

\section{Introduction and calculation}

 The lattice numerical calculation of the matrix element of any
operator $\cal{A}$ with mass dimension $d$ yields $a^d\cal{A}$, where 
$a$ is the lattice spacing, up to scaling violating contributions of 
order $O(a^{d+1})$. The calculation of $a$ as a function of the coupling 
constant can be improved
if an effective scheme is used \cite{parisi1,parisi2}. A nice example of such
improvement has been shown to occur in pure Yang-Mills
theories \cite{npb} and is known to work rather well in other scalar theories
(see for example \cite{wolff,abc} for the 2D spin models). 

 In this paper we calculate the 3-loop internal energy in full QCD with
Wilson fermions on the lattice. The 3-loop beta function for this theory
has been recently calculated \cite{harisettore,proc97} and we can compute the
corresponding corrections to asymptotic scaling in both the standard 
and effective schemes. 

 We will use the Wilson action $S$ with $N_f$ flavours and gauge group
$SU(N)$. This action is 
\begin{eqnarray}
  S &=& S_W + S_f , \nonumber \\
 S_W &=& \beta \sum_\Box E_W(\Box ) , \\
 S_f &=& \sum_x E_f(x) \nonumber
\label{Swilson}
\end{eqnarray}
with
\begin{eqnarray}
 E_W(\Box ) &=&  1 - {1\over N} \hbox{Re Tr} \left( \Box \right) , \nonumber \\
 E_f(x) &=& \sum_{\rm flavours} 
         a^4 \Bigg[ \left(m + {4 r \over a} \right)
         \overline\psi_x \psi_x -  \\ 
      & & {1 \over {2 a}} \sum_\mu
         \left( \overline\psi_{x+\hat\mu} \left( r +
         \gamma_\mu \right) U^\dagger_\mu(x) \psi_x +
         \overline\psi_x \left(r - \gamma_\mu \right)
         U_\mu(x) \psi_{x+\hat\mu} \right) \Bigg] . \nonumber
\label{energy}
\end{eqnarray}
$\beta$ is the coupling on the lattice $\beta\equiv 2N/g^2$ and
$r$ the Wilson parameter.
$\Box$~stands for the plaquette and $U_\mu(x)\equiv \exp
\left(i a g A_\mu(x)\right)$ for the link at the
site $x$ pointing towards $x +\hat\mu$.
We will parametrize the results in terms of $N$, $N_f$ and 
the couple $(\kappa,r)$ where $\kappa$ is the usual hopping parameter
\begin{equation} 
 \kappa = { 1  \over {8 r + 2 am}}.
\label{kappa} 
\end{equation}

 The average of $E_f$ can be straightforwardly computed to all orders
by rescaling the fermionic action $S_f$ by a factor $\epsilon$
under the fermionic path integral $Z^f$
\begin{equation}
 Z^f(\epsilon) \equiv \int {\cal{D}}\overline\psi(x) 
               {\cal{D}}\psi(x) \exp\left(
               - \epsilon S_f \right) = \epsilon^{4 V N N_f} 
               Z^f(\epsilon =1)
\label{Zf}
\end{equation}
and by using 
\begin{equation}
 \langle E_f \rangle = -  {\partial \over {\partial \epsilon}}
                       \left({{\ln Z^f(\epsilon)} \over V}
                       \right)_{\epsilon =1} = - 4 N N_f.
\label{ef}
\end{equation}

On the other hand, the average of $E_W$ in presence of the 
action Eq.(\ref{Swilson}) is calculated in perturbation theory
\begin{equation}
 \langle E_W \rangle = c_1 \; g^2 + c_2 \; g^4 + c_3 \; g^6 + \cdots
\label{expansion}
\end{equation}
The $n$-loop coefficient can be written as $c_n = c^g_n + c^f_n$ where 
$c^g_n$ is the pure Yang-Mills contribution, known 
since ref.~\cite{gcrossi,c3feo}
up to 3-loops, and $c_n^f$ is the fermionic contribution.
To calculate $c_n^f$ we will first compute the free energy
$-(\ln Z)/V$ up to 3~loops, $Z$ being the full partition function
\begin{equation}
 Z \equiv \int {\cal D}U_\mu(x) {\cal D}\overline\psi(x) 
          {\cal D}\psi(x) \exp(-S) .
\label{Z}
\end{equation}
The average of $E_W$ is then extracted as follows
\begin{equation}
 \langle E_W \rangle = - {1 \over 6}\, {\partial \over {\partial \beta}}\,
                       \left( {\ln Z \over V} \right) .
\label{e}
\end{equation}

The Feynman diagrams necessary 
to calculate the fermion contribution to the 
free energy up to 3~loops are shown in Fig.~1.
We have used the Feynman gauge. The involved algebra of the lattice
perturbation theory was carried out by making use of the computer code
developed by us~\cite{npbus}.

After grouping the diagrams in several infrared-finite sets, we 
calculated the resulting finite integrals on finite lattice-sizes $L$ and then
the results were extrapolated to infinite size.
The extrapolating function was of the type~\cite{luscher,caracc2}
\begin{equation}
 a_0 + \sum_{i \leq 2j ,\; j=1,2,\cdots}
 a_{ij} {{\left(\ln L\right)^i } \over L^{2j}}.
\label{extrapolation}
\end{equation}
We used a broad spectrum of such functional forms and analyzed the
quality of each extrapolation to assign
a weight to each one, to finally produce a reliable
estimate of the systematic error. 
The criterion to judge the quality of the extrapolation was 
based on the accuracy of the 
the fitted functional form to reproduce known results at finite 
lattice-sizes.
The different functional forms were
obtained by truncating the series in Eq.(\ref{extrapolation}) at
different values of $j$ and assuming vanishing coefficients $a_{ij}$ 
for some $i$ and $j$.

Recall \cite{gcrossi,c3feo} that the pure gluonic contributions 
are~(presented here with improved accuracy)
\begin{eqnarray}
 c_1^g &=& {{N^2 - 1} \over 8 \;N} , \nonumber \\
 c_2^g &=& \left( N^2 - 1 \right) \left(0.0051069297 -
           {1 \over {128 \; N^2}} \right) , \\
 c_3^g &=& \left( N^2 - 1 \right) \Bigg( {0.0023152583(50) \over N^3}  -
        {0.002265487(17) \over N} + \nonumber \\
       & & \; \; \; \;\;\;\;\;\;\;\;\;\;\;\;\;\;\;\;\;
         0.000794223(19) \; N \Bigg) . \nonumber 
\label{cg}
\end{eqnarray}
The 2-loop coefficient can be written in closed form as
\begin{equation}
 \left( N^2 - 1 \right) \left( \
{1\over {384}} + {{{ P_1}}\over {24}} - {{{{{ P_1}}^2}}\over {12}} - 
   {{{ Q_1}}\over {24}} - {{{ Q_2}}\over {288}} - {1 \over {128 \; N^2}}
\right),
\end{equation}
where $P_1$, $Q_1$ and $Q_2$ are finite integrals defined and evaluated
in~\cite{lusweisz}.


In tables~I and~II we show $c_2^f$ and $c_3^f$ for several pairs
$(\kappa,r)$. They are parametrized in terms of four constants
$h_2$, $h_{30}$, $h_{31}$ and $h_{32}$ as follows
\begin{eqnarray}
 c_1^f &=& 0  \; ,\nonumber \\
 c_2^f &=&\left(N^2 - 1\right) h_2 \; {{ N_f} \over N} , \\
 c_3^f &=& \left(N^2 - 1\right)
          \left( h_{30} \; N_f + h_{31} \; {N_f \over N^2} + 
           h_{32} \; {N_f^2 \over N} \right) . \nonumber
\label{cf}
\end{eqnarray}

In table~III we show the result for $c_2=c_2^g + c_2^f$ and 
$c_3=c_3^g + c_3^f$ for $r=1$, $N_f=3$
and $N=2$. In table~IV the result for $r=1$, $N_f=3$ and $N=3$ is shown.

\section{Corrections to asymptotic scaling}

The beta function in QCD with $N_f$ Wilson fermions can be written as
\begin{equation}
 {\overline\beta}^L(g) \equiv - a {{d g} \over {d a}}|_{g_R,\;\mu} = 
                    -b_0 g^3 - b_1 g^5 - b_2 g^7 - \cdots
\label{beta}
\end{equation}
The non-universal 3-loop coefficient $b_2$ has been recently
calculated in ref.~\cite{harisettore}, (see also~\cite{proc97}).
In terms of the bare coupling $g$, the lattice spacing $a$ approaches
the continuum limit as
\begin{equation}
 a\Lambda_L = \exp\left( - {1 \over {2 b_0 g^2}} \right) 
              \left( b_0 g^2 \right)^{- b_1 / 2 b_0^2} 
              \left( 1 + q \; g^2 +\cdots \right) ,
\label{ascaling}
\end{equation}
where $q$ is the 3-loop correction to asymptotic scaling
\begin{equation}
 q \equiv  {{b_1^2 - b_2 b_0} \over {2b_0^3}} .
\end{equation}

Other couplings can be defined. 
A popular effective coupling in terms of the plaquette
energy is~\cite{parisi1,parisi2}
\begin{equation}
 g_{E_W}^2 \equiv {1 \over c_1} \; \langle E_W \rangle_{MC} ,
\label{ge}
\end{equation}
where $\langle \cdot \rangle_{MC}$ indicates Monte Carlo average.
In terms of $g_{E_W}$ the approach of the lattice spacing to the
continuum limit is written as in Eq.(\ref{ascaling}) with $q_{E_W}$
instead of $q$. This 3-loop correction to asymptotic scaling is
\begin{equation}
q_{E_W} = q - {{b_0 \, c_3 - b_1 \, c_2 - b_0 \, c_2^2/c_1} \over 
{2 \, c_1 \, b_0^2}} .
\label{qe}
\end{equation}
Moreover the lattice Lambda parameters are related by the equation
\begin{equation}
 \Lambda_{E_W} = \exp \left( {c_2 \over {2\, b_0 \, c_1}} \right) \;
 \Lambda_L .
\label{lambda}
\end{equation}

We give the coefficients $q_{E_W}$ for $SU(2)$ in table~V and
for $SU(3)$ in table~VI. In both cases we show the result for
$r=1$ and for several choices of $\kappa$. 
These results must be compared with the value of $q$ in the standard
bare coupling scheme. This is~\cite{harisettore}
\begin{eqnarray}
SU(2) \;\; \rightarrow & & q(N_f =1)=0.12617(4) \;\;\;\; q(N_f =3)=0.2547(1) 
     ,  \nonumber \\
SU(3) \;\; \rightarrow & & q(N_f =1)=0.23956(4) \;\;\;\; q(N_f =3)=0.3681(2) ,
\label{q}
\end{eqnarray}
(notice that in the standard bare coupling scheme, $q$ does not 
depend on the fermion mass~\cite{harisettore}).
The 3-loop correction to asymptotic scaling for $N_f=3$, $N=3$
is $\sim 37\%$ in the standard scheme and $\sim 18-21\%$ in 
 the~$\langle E_W \rangle$ scheme. Apparently this improvement
is not as dramatic as it was for the pure gluonic case \cite{npb}
(for $SU(3)$ $q\sim 19\%$ in the standard scheme and $q\sim 1\%$ in
the energy scheme). The improvement therefore seems to be more
efficient in the quenched case. This fact can also be seen from tables~V
and~VI where the approach to the quenched case (either
$N_f \longrightarrow 0$ or $\kappa \longrightarrow 0$) is 
accompained by the lowering of $q_{E_W}$. 
However, the only relevant test for the 
improvement in asymptotic scaling from
the standard to some energy scheme is the practical use of the effective
scheme. For example, the 2D spin models
\cite{wolff,abc} show a definite improvement in spite of the
behaviour of the 3-loop coefficient $q$ which for the $O(3)$ 
models passes from $q=-0.09138$ in the standard scheme to~$0.1694$
in the energy scheme.


\section{Discussion}

We have calculated the free energy up to 3~loops in QCD with
Wilson fermions. We have given the result of the internal energy
average $\langle E_W\rangle$ 
as a function of the number of fermions $N_f$, their masses and the
Wilson parameter $r$. These expansions have been used to study
the 3-loop corrections to asymptotic scaling in this theory. We have
found that at 3~loops the corresponding energy scheme \cite{parisi1,parisi2} 
provides a moderate improvement with respect to the standard scheme.
However, only the practical use of this scheme in particular problems
will reveal how useful it is. 

 We can numerically compute the set of coefficients $c_n^f$ 
and $q$ also for other choices of $(\kappa, r)$, $N_f$ and $N$.

\section{Acknowledgements}
We thank CNUCE (Pisa) for qualified technical 
assistance in the use of their IBM--SP2.
We would also like to thank Andrea Pelissetto and Ettore Vicari for 
useful discussions.

\newpage


\newpage

\noindent{\bf Figure caption}

\begin{enumerate}

\item[Figure 1.] Feynman diagrams contributing to the fermionic
sector of $\ln Z/V$ at 2 and 3~loops. Curly, arrowed and dashed lines
indicate gluons, fermions and ghosts respectively.

\end{enumerate}

\vskip 2cm

\noindent{\bf Table captions}

\vskip 5mm

\begin{enumerate}

\item[Table I.] 2-loop fermion contribution to the internal energy 
$\langle E_W \rangle$. See text for notation.

\item[Table II.] 3-loop fermion contribution to the internal energy 
$\langle E_W \rangle$. See text for notation.

\item[Table III.] Coefficients $c_2$ and $c_3$ for $SU(2)$ and
$N_f=3$ at $r=1$.

\item[Table IV.] Coefficients $c_2$ and $c_3$ for $SU(3)$ and
$N_f=3$ at $r=1$.

\item[Table V.] 3-loop correction to asymptotic scaling
in $SU(2)$ with $r=1$ in the energy scheme $\langle E_W \rangle$.

\item[Table VI.] 3-loop correction to asymptotic scaling
in $SU(3)$ with $r=1$ in the energy scheme $\langle E_W \rangle$.


\end{enumerate}

\newpage

\centerline{\bf Table I}

\vskip 5mm

\moveright 1.2 in
\vbox{\offinterlineskip
\halign{\strut
\vrule \hfil\quad $#$ \hfil \quad &
\vrule \hfil\quad $#$ \hfil \quad &
\vrule \hfil\quad $#$ \hfil \quad \vrule \cr
\noalign{\hrule}
\kappa &
r &
h_2 \times 10^3\cr
\noalign{\hrule}
0.1675  &  0.1  &  -0.269079(27) \cr
\noalign{\hrule}
0.161  &  0.1  &  -0.236105(26) \cr
\noalign{\hrule}
0.156  &  0.1  &  -0.212480(25) \cr
\noalign{\hrule}
0.1675  &  0.5  &  -0.502547(23) \cr
\noalign{\hrule}
0.161  &  0.5  &  -0.434010(22) \cr
\noalign{\hrule}
0.156  &  0.5  &  -0.385777(21) \cr
\noalign{\hrule}
0.1675  &  1  &  -2.408379(14) \cr
\noalign{\hrule}
0.164  &  1  &  -2.273850(12) \cr
\noalign{\hrule}
0.16  &  1  &  -2.117310(36) \cr
\noalign{\hrule}
0.1575  &  1  &  -2.017993(24) \cr
\noalign{\hrule}
0.156  &  1  &  -1.957882(14) \cr
\noalign{\hrule}
0.1675  &  2  &  -6.44013(90) \cr
\noalign{\hrule}
0.161  &  2  &  -6.29318(41) \cr
\noalign{\hrule}
0.156  &  2  &  -6.16472(95) \cr
\noalign{\hrule}
}}

\vskip 1cm

\centerline{\bf Table II}

\vskip 5mm

\moveright -.4 in
\vbox{\offinterlineskip
\halign{\strut
\vrule \hfil\quad $#$ \hfil \quad &
\vrule \hfil\quad $#$ \hfil \quad &
\vrule \hfil\quad $#$ \hfil \quad &
\vrule \hfil\quad $#$ \hfil \quad &
\vrule \hfil\quad $#$ \hfil \quad \vrule \cr
\noalign{\hrule}
\kappa &
r &
h_{30} \times 10^3 &
h_{31} \times 10^3 &
h_{32} \times 10^3 \cr
\noalign{\hrule}
0.1675  &  0.1  &
-0.0194886(93)   &
0.030311(54) &
0.000585(20) \cr
\noalign{\hrule}
0.161  &  0.1  &
-0.017144(10) &
0.026733(47) &
0.000451(18) \cr
\noalign{\hrule}
0.156  &  0.1  &
-0.015466(11) &
0.024140(41) &
0.000366(17) \cr
\noalign{\hrule}
0.1675  &  0.5  &
-0.032691(24) &
0.050837(49) &
0.002041(15) \cr
\noalign{\hrule}
0.161  &  0.5  &
-0.028269(21) &
0.044149(52) &
0.001521(16) \cr
\noalign{\hrule}
0.156  &  0.5  &
-0.025205(15) &
0.039464(51) &
0.001202(14) \cr
\noalign{\hrule}
0.1675  &  1  &
-0.361568(64) &
0.4271834(82) &
0.049019(10) \cr
\noalign{\hrule}
0.164  &  1  &
-0.330133(54) &
0.390773(14) &
0.043856(15) \cr
\noalign{\hrule}
0.16  &  1  &
-0.293659(46) &
0.348654(20) &
0.038177(14) \cr
\noalign{\hrule}
0.1575  &  1  &
-0.270560(43) &
0.322111(22) &
0.034772(11) \cr
\noalign{\hrule}
0.156  &  1  &
-0.256620(43) &
0.306132(22) &
0.0327854(66) \cr
\noalign{\hrule}
0.1675  &  2  &
-1.4249(32) &
1.6754(19) &
0.34483(95) \cr
\noalign{\hrule}
0.161  &  2  &
-1.3838(44) &
1.6257(34) &
0.3297(14) \cr
\noalign{\hrule}
0.156  &  2  &
-1.3541(29) &
1.5868(22) &
0.3166(14) \cr
\noalign{\hrule}
}}

\newpage

\centerline{\bf Table III}

\vskip 5mm

\moveright 1.2 in
\vbox{\offinterlineskip
\halign{\strut
\vrule \hfil\quad $#$ \hfil \quad &
\vrule \hfil\quad $#$ \hfil \quad &
\vrule \hfil\quad $#$ \hfil \quad \vrule \cr
\noalign{\hrule}
\kappa &
c_2 \times 10^3&
c_3 \times 10^3\cr
\noalign{\hrule}
0.1675 & -1.376291(62) & 0.60413(60) \cr
\noalign{\hrule}
0.164 & -0.770912(55)   & 0.73543(54) \cr
\noalign{\hrule}
0.16  & -0.06648(16) & 0.89226(47) \cr
\noalign{\hrule}
0.1575 & 0.38045(11) &  0.99446(44) \cr
\noalign{\hrule}
0.156 & 0.650945(61) & 1.05715(42) \cr
\noalign{\hrule}
}}

\vskip 1cm

\centerline{\bf Table IV}

\vskip 5mm

\moveright 1.2 in
\vbox{\offinterlineskip
\halign{\strut
\vrule \hfil\quad $#$ \hfil \quad &
\vrule \hfil\quad $#$ \hfil \quad &
\vrule \hfil\quad $#$ \hfil \quad \vrule \cr
\noalign{\hrule}
\kappa &
c_2 \times 10^3&
c_3 \times 10^3\cr
\noalign{\hrule}
0.1675 & 14.64396(11) & 7.3440(16) \cr
\noalign{\hrule}
0.164 & 15.720192(98) & 7.8775(14) \cr
\noalign{\hrule}
0.16 & 16.97251(29)  & 8.5042(12) \cr
\noalign{\hrule}
0.1575  & 17.76705(19)  & 8.9061(12) \cr
\noalign{\hrule}
0.156  &  18.24794(11)  &  9.1504(12) \cr
\noalign{\hrule}
}}

\vskip 1cm

\newpage

\centerline{\bf Table V}

\vskip 5mm

\moveright 1.2 in
\vbox{\offinterlineskip
\halign{\strut
\vrule \hfil\quad $#$ \hfil \quad & 
\vrule \hfil\quad $#$ \hfil \quad & 
\vrule \hfil\quad $#$ \hfil \quad \vrule \cr
\noalign{\hrule}
\kappa &
q_{E_W}(N_f = 1) &
q_{E_W}(N_f = 3) \cr
\noalign{\hrule}
0.1675 & 0.05318(5) & 0.2052(2) \cr
\noalign{\hrule}
0.164 & 0.05069(4)  & 0.1954(1) \cr
\noalign{\hrule}
0.16 & 0.04782(5) & 0.1842(2) \cr
\noalign{\hrule}
0.1575 & 0.04599(5) & 0.1770(1) \cr
\noalign{\hrule}
0.156 & 0.04489(5) & 0.1727(1) \cr
\noalign{\hrule}
}}

\vskip 1cm

\centerline{\bf Table VI}

\vskip 5mm

\moveright 1.2 in
\vbox{\offinterlineskip
\halign{\strut
\vrule \hfil\quad $#$ \hfil \quad & 
\vrule \hfil\quad $#$ \hfil \quad & 
\vrule \hfil\quad $#$ \hfil \quad \vrule \cr
\noalign{\hrule}
\kappa &
q_{E_W}(N_f = 1) &
q_{E_W}(N_f = 3) \cr
\noalign{\hrule}
0.1675 & 0.06645(4) & 0.2091(2) \cr
\noalign{\hrule}
0.164 & 0.06355(4) & 0.1989(2) \cr
\noalign{\hrule}
0.16 & 0.06021(4) & 0.1871(1) \cr
\noalign{\hrule}
0.1575 & 0.05809(4) &  0.1796(1) \cr
\noalign{\hrule}
0.156 & 0.05681(4) &  0.1752(2) \cr
\noalign{\hrule}
}}






\begin{thebibliography}{99}
\bibitem{parisi1} G. Parisi, in ``High Energy Physics'', 1980,
 Proceedings of the XXth International AIP Conference, ed.
 L. Durand and L. G. Pondrom, p. 1531.
\bibitem{parisi2} G. Martinelli, G. Parisi, R. Petronzio, 
 Phys. Lett. B100 (1981) 485.
\bibitem{npb} B. All\'es, A. Feo, H. Panagopoulos, Nucl. Phys. B491 (1997) 498.
\bibitem{wolff} U. Wolff, Phys. Lett. B248 (1990) 335.
\bibitem{abc} B. All\'es, A. Buonanno, G. Cella, Nucl. Phys. B500 (1997) 513.
\bibitem{harisettore} C. Christou, A. Feo, H. Panagopoulos, E. Vicari,
 preprint IFUP-TH 65/97.
\bibitem{proc97} B. All\'es, C. Christou, A. Feo, H. Panagopoulos,
 E. Vicari, hep-lat/9710018, to appear in Proceedings XVth International
 Symposium on Lattice Field Theory, Lattice-1997.
\bibitem{gcrossi} A. Di Giacomo, G. C. Rossi, Phys. Lett. B100 (1981) 481.
\bibitem{c3feo} B. All\'es, M. Campostrini, A. Feo, H. Panagopoulos,
 Phys. Lett. B324 (1994) 433.
\bibitem{npbus} B. All\'es, M. Campostrini, A. Feo, H. Panagopoulos,
 Nucl. Phys. B413 (1994) 553.
\bibitem{luscher} M. L\"uscher, P. Weisz, Nucl. Phys. B266 (1986) 309.
\bibitem{caracc2} S. Caracciolo, A. Pelissetto, in preparation.
\bibitem{lusweisz} M. L\"uscher, P. Weisz, Nucl. Phys. B445 (1995) 429.
\end{thebibliography}
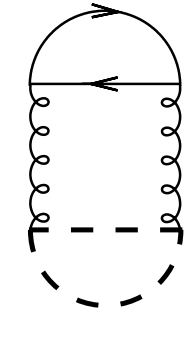
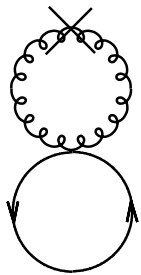
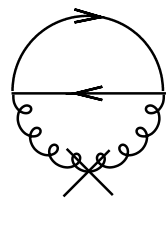
\end{document}